\documentstyle{article}

\begin{document}

{\bf TRANSFORMATION OF THERMAL ENERGY IN ELECTRIC ENERGY IN AN
INHOMOGENEOUS SUPERCONDUCTING RING}

\

{\bf A.V.Nikulov}

Institute of Microelectronics Technology and High Purity
Materials, Russian Academy of Sciences, 142432 Chernogolovka, Moscow
Region, Russia.

\begin{abstract}
An inhomogeneous superconducting ring (hollow cylinder) placed in a
magnetic field is considered. It is shown that a direct voltage appears on
section with lowest critical temperature when it is switched periodically
from the normal state in the superconducting state and backwards, if the
magnetic flux contained within the ring is not divisible by the flux
quantum. The superconducting transition can be first order in this case. In
the vicinity of this transition, thermal fluctuations can induce the
voltage in the ring with rather small sizes. \end{abstract}

\vspace{0.3in}

\section{Introduction}

	Superconductivity is a macroscopic quantum phenomenon. One of the
consequences of this is the periodical dependence of energy of a
superconducting ring on a magnetic flux within this ring. This
dependence is caused by a quantization of the velocity of the
superconducting electrons $v_{s}$. According to (See ref.\cite{tink75})

$$v_{s} = \frac{1}{m}(\hbar \frac{d\phi}{dr} - \frac{2e}{c}A) =
\frac{2e}{mc}(\frac{\Phi_{0}}{2\pi }\frac{d\phi}{dr} - A) \eqno{(1)}$$
the velocity along the ring (or tube) circumference must have fixed values
dependent on the magnetic flux because

$$\int_{l} dl v_{s} = \frac{2e}{mc}(\Phi_{0}n - \Phi) \eqno{(2)}$$
and $n = \int_{l} dl(1/2\pi )d\phi/dr$ must be an integer number since the
wave function $\Psi = |\Psi | \exp(i\phi)$ of the superconducting electrons
must be a simple function. Where $\phi$ is the phase of the wave function;
$\Phi_{0} = \pi \hbar c/e =2.07 \ 10^{-7} G \ cm^{2}$ is the flux quantum;
A is the vector potential; m is the electron mass and e is the electron
charge; $l=2\pi R$ is the ring circumference; R is the ring radius; $\Phi =
\int_{l}dlA$ is the magnetic flux contained within the ring.

	The energy of the superconductor increases with the superconducting
electron velocity increasing. Therefore the $|v_{s}|$ tends towards a
minimum possible value. If $\Phi/\Phi_{0}$ is an integer number, the
velocity is equal to zero. But $v_{s}$ cannot be equal to zero if
$\Phi/\Phi_{0}$ is not an integer number. Consequently, the energy of the
superconducting state of the ring depends in a periodic manner on the
magnetic field value. The energy of the supeconducting ring changes by two
reason: because the kinetic energy of the superconducting electrons
$n_{s}slmv_{s}^{2}/2$ changes and because the energy of a magnetic field
$LI_{s}^{2}/2$ induced by superconducting current $I_{s} = sj_{s} =
s2en_{s}$ changes. Here L is the inductance of the ring; $n_{s}$ is the
superconducting pair density; s is the area of cross-section of the ring
wall.

	The first is the cause of the Little-Parks effect \cite{little}. Little
and Parks discovered that the critical temperature, $T_{c}$, of a
superconducting tube with narrow wall depends in a periodic way on a
magnetic flux value within the tube. This effect has been explained by
M.Tinkham \cite{tinkham}. According to ref.\cite{tinkham}, the critical
temperature of the homogeneous ring (which we considered now) is shifted
periodically in the magnetic field:

$$T_{c}(\Phi) = T_{c}[1 - (\xi(0)/R)^{2}(n-\Phi /\Phi_{0})^{2}] \eqno{(3)}
$$ because the kinetic energy changes periodically with magnetic field.
Here $\xi(0)$ is the coherence length at T = 0. The value of  ($n-\Phi
/\Phi_{0}$) changes from -0.5 to 0.5. The $T_{c}$ shift is visible if the
tube radius is small enough (if $R \simeq \xi(0)$).

	The magnetic field energy $F_{L} = LI_{s}^{2}/2 =
Ls^{2}2e^{2}v_{s}^{2}n_{s}^{2}$ does not influence on the critical
temperature value because it is proportional to $n_{s}^{2}$. A value of
this energy depends on the temperature because the $n_{s}$ value depends
on the temperature. The $n_{s}$ value change causes the
superconducting current change and a voltage in consequence with
electromagnetic induction law. Consequently the superconducting ring can
be used for transformation the thermal energy into the electric energy if
the velocity $v_{s} \neq 0$. A temperature change will induce a voltage in
the ring if $v_{s} \neq 0$. In the present paper a most interesting case -
a induction of direct voltage in an inhomogeneous ring - is considered.

	A ring (tube) with the narrow wall (the wall thickness $w \ll R,
\lambda$) is considered. In this case $\Phi = BS \simeq HS$, because the
magnetic field induced by the superconducting current in the ring is small.
Here H is the magnetic field induced by an external magnet, $S=\pi R^{2}$
is the ring area and $\lambda$ is the penetration depth of the magnetic
field. We consider a ring whose critical temperature varies along the
circumference $l = 2\pi R$, but is constant along the height h. In such a
ring, the magnetic flux shifts the critical temperature of a section with a
lowest $T_{c}$ value only. When the superconducting state is closed in the
ring, the current of the superconducting electrons must appear as a
consequence of the relation (2) if $\Phi/\Phi_{0}$ is not an integer
number. Therefore the lowest $T_{c}$ value will be shifted periodically in
the magnetic field as well as $T_{c}$ of the homogeneous ring.

\section{Inhomogeneous superconducting ring as a thermal-electric machine
of direct current}

	It is obvious that the current value must be constant along the
circumference, because the capacitance is small. The value of the
superconductor current must be constant if the current of the normal
electrons is absent. Therefore the velocity of the superconducting
electrons can not be a constant value along the circumference of a
inhomogeneous ring if the superconducting pair density is not constant. (I
consider the ring with identical areas along the circumference.) Let us
consider a ring consisting of two sections $l_{a}$ and $l_{b}$
($l_{a}+l_{b} = l =2\pi R$) with different values of the critical
temperature $T_{ca} > T_{cb}$. According to the relation (2) the
superconducting current along the ring circumference, $I_{s}$, must appear
below $T_{cb}$ if $\Phi/\Phi_{0}$ is not an integer number. Then

$$I_{s}=I_{sa}=s_{a}j_{sa}=s_{a}2en_{sa}v_{sa}=I_{sb}=s_{b}j_{sb}=
s_{b}2en_{sb}v_{sb} \eqno{(4)} $$
if the normal current is absent. Here $n_{sa}$ and $n_{sb}$ are the
densities of the superconducting pair in the sections $l_{a}$ and $l_{b}$;
$v_{sa}$ and $v_{sb}$ are the velocities of the superconducting pairs in
the sections $l_{a}$ and $l_{b}$ and $s_{a}$ and $s_{b}$ are the areas of
wall section of $l_{a}$ and $l_{b}$. $s = s_{a} = s_{b} = wh$.  $\int_{l}
dl v_{s} = v_{sa}l_{a} + v_{sb}l_{b}$. Therefore according to (2) and (4)

$$v_{sa} = \frac{2e}{mc}\frac{n_{sb}}{(l_{a}n_{sb}+l_{b}n_{sa})}
(\Phi_{o}n-\Phi); \ v_{sb} = \frac{2e}{mc}
\frac{n_{sa}}{(l_{a}n_{sb}+l_{b}n_{sa})} (\Phi_{o}n-\Phi) \eqno{(5)}$$

$$I_{s} = \frac{s4e^{2}}{mc}
\frac{n_{sa}n_{sb}}{(l_{a}n_{sb}+l_{b}n_{sa})} (\Phi_{o}n-\Phi)
\eqno{(5a)}$$

	According to (5a) a change of the superconducting pair density
induces a change of the superconducting current if $\Phi_{o}n-\Phi \neq
0$. The change of the superconducting current induces the change of the
magnetic flux $\Phi = H\pi R^{2} + L(I_{s}+I_{n})$ and, as a consequence,
induces the voltage and the current of the normal electrons (the normal
current, $I_{n}$). The total current $I=I_{s}+I_{n}$ must be equal in both
sections, because the capacitance is small. But $I_{sa}$ can be no equal to
$I_{sb}$. Then $I_{na} \neq I_{nb}$ and consequently, the
potential difference $dU/dl$ exists along the ring circumference. The
electric field along the ring circumference $E(r)$ is equal to

$$E(r) = -\frac{dU}{dl} -\frac{1}{l}\frac{d\Phi}{dt} =-\frac{dU}{dl} -
\frac{L}{l}\frac{dI}{dt}= \frac{\rho_{n}}{s} I_{n} \eqno{(6)}$$
where $\rho_{n}$ is the normal resistivity.

	Because the normal current exists the relations (5) becomes no valid.
The velocity $v_{sa}$ and the current $I_{ca}$ can be not equal zero even at
$n_{sb} = 0$. The current decreases during the decay time of the normal
current $L/R_{nb}$. $R_{bn} = \rho_{bn}l_{b}/s$ is the resistance of the
section $l_{b}$ in the normal state.

	The voltage, the normal current, and superconducting current change
periodically in time if the $n_{sb}$ value changes periodically. But in
addition a direct potential difference $U_{b}$ can appear if the $l_{b}$
section is switched from the normal state in the superconducting state
and backwards i.e. some times ($t_{n}$) $n_{sb} = 0$ and some times
($t_{s}$) $n_{sb} \neq 0$. Let us consider two limit cases:
$t_{n} \ll L/R_{nb}$ and $t_{n} \gg L/R_{nb}$. $t_{s} \gg L/R_{nb}$ and
$l_{a} \gg l_{b}$ in the both cases.

	In the first case the total current I is approximately constant in time
and because $t_{s} \gg L/R_{bn}$

$$I \simeq s2en_{sa}<v_{sa}> \simeq s\frac{4e^{2}}{mc}\frac{n_{sa}<n_{sb}>}
{l_{b}n_{sa}+l_{a}<n_{sb}>}(\Phi_{0}n-\Phi) \eqno{(7)}$$
Here $<n_{sb}>$ is a average value of $n_{sb}$.  The average value of the
resistivity of the $l_{b}$ section, $\rho_{b} \simeq
\rho_{bn}t_{n}/(t_{s}+t_{n})$. Consequently,

$$U_{b} = R_{b}I \simeq \frac{l_{b}<n_{sb}>} {l_{b}n_{sa}+l_{a}<n_{sb}>}
\frac{(\Phi_{0}n-\Phi)}{\lambda_{La}^{2}}\rho_{b} \eqno{(8)}$$
where $\lambda_{La} = (mc/4e^{2}n_{sa})^{1/2}$ is the London penetration
depth. According to (8) $U_{b} \neq 0$ at $<n_{sb}> \neq 0$ and $\rho_{b}
\neq 0$. Consequently, the direct potential difference can be observed if
we change the temperature inside the region of the resistive transition  of
the section $l_{b}$ (where $0 < \rho_{b} < \rho_{bn}$).

	In the second case

$$U_{b} \simeq \frac{s<n_{sb}>} {l_{b}n_{sa}+l_{a}<n_{sb}>}
\frac{(\Phi_{0}n-\Phi)}{\lambda_{La}^{2}}Lf \eqno{(9)}$$
where f is the frequency of the switching from the normal into the
superconducting state.

	Thus, the inhomogeneous superconducting ring can be used as a
thermal-electric machine of direct current. The power of this machine is
small. It decreases with the increasing of the ring radius. Let us to
evaluate the maximum power at $l_{b}n_{sa} \ll l_{a}<n_{sb}>$. This
condition means that the temperature of $l_{b}$ changes enough strongly.
The power will be maximum at $t_{n} \simeq t_{s} \simeq L/R_{bn}$. The
frequency of the switching $f = R_{bn}/2L$ in this case. The power of the
ring can not exceed

$$W = IU_{b} = \frac{sl_{b}\rho_{bn}\Phi_{0}^{2}}{2l_{a}^{2}
\lambda_{La}^{4}} (n-\Phi/\Phi_{0})^{2} < \frac{sl_{b}
\rho_{bn}\Phi_{0}^{2}} {8l_{a}^{2}\lambda_{La}^{4}} \eqno{(10)}$$
For example, at $s = 0.1 \mu m$, $l_{b} = 0.1 \mu m$, $l_{a} = 1 \mu m$,
$\lambda_{La} = 0.1 \mu m$, $\rho_{bn} = 100 \mu \Omega \ cm$ the power is
smaller than $W = 10^{-4} Vt$. The power can be increased by the ring
height h increasing.

	Above we considered the case when the changes of the temperature is
enough strong. In the opposite case, when the temperature change is small,
$l_{b}n_{sa} \gg l_{a}<n_{sb}>$ and therefore the current and the voltage
are proportional to the $<n_{sb}>/n_{sa}$ value. Because $n_{sb}$ must be
equal zero some times, this means that the voltage is proportional to the
amplitude of the temperature change. Consequently, the inhomogeneous ring
is a classical thermal machine with a maximum efficiency in the Carno cycle
\cite{kittel}.

	At a small $l_{b}/l_{a}$ value and a enough big $|(n-\Phi/\Phi_{0})|$
value the superconducting transition of $l_{b}$ is first order. In this
case in order to switch the $l_{b}$ section from the normal state in the
superconducting state and backwards the temperature must be change on a
finite value, because the hysteresis of the superconducting transition
exists.

\section{First order superconducting phase transition}

	According to (5) the $v_{sb}$ value decreases with the $n_{sb}$ value
increasing. Therefore the dependence of the energy of the superconducting
state on the $n_{sb}$ value can have a maximum in some temperature region
at $T \simeq T_{cb}(\Phi)$. The presence of such a maximum means that the
superconducting transition is a first order phase transition.

	The existence of the maximum and the width of the temperature region
where the maximum exists depends on the $n-\Phi/\Phi_{0}$ value and on the
ring parameters: $l_{a}$, $l_{b}$, w, h and $T_{ca}/T_{cb}$. These
dependencies can be reduced to two parameters, $B_{f}$ and $L_{I}$, which
are introduced below. It is obvious that the maximum can exist at only
$n-\Phi/\Phi_{0} \neq 0$. Therefore only this case is considered
below.

	The Ginsburg-Landau free energy of the ring can be written as

$$F_{GL} = s[l_{a}((\alpha_{a}+\frac{mv_{sa}^{2}}{2})n_{sa}+
\frac{\beta_{a}}{2}n_{sa}^{2})
+ l_{b}((\alpha_{b}+\frac{mv_{sb}^{2}}{2})n_{sb}+ \frac{\beta_{b}}{2}
n_{sb}^{2})] + \frac{LI_{s}^{2}}{2} \eqno{(11)}$$
Here $\alpha_{a} = \alpha_{a0}(T/T_{ca}-1)$, $\beta_{a}$, $\alpha_{b} =
\alpha_{b0}(T/T_{cb}-1)$ and $\beta_{b}$ are the coefficients of the
Ginsburg-Landau theory. We do not consider the energy connected with the
density gradient of the superconducting pair. It can be shown that this
does not influence essentially the results obtained below.

	The Ginsburg-Landau free energy (11) consists of $F_{GL,la}$ (the energy
of the section $l_{a}$), $F_{GL,lb}$ (the energy of the section $l_{a}$)
and $F_{L}$ (the energy of the magnetic field induced by the
superconducting current):

$$F_{GL} = F_{GL,la} + F_{GL,lb} + F_{L} \eqno{(12)}$$

Substituting the relation (4) for the superconducting current and the
relation (5) for the velocity of the superconducting electrons into the
relation (11), we obtain

$$F_{GL,la} = sl_{a}(\alpha_{a}(\Phi ,n_{sa},n_{sb})n_{sa} +
\frac{\beta_{a}}{2}n_{sa}^{2}) \eqno{(12a)}$$

$$F_{GL,lb} = sl_{b}(\alpha_{b}(\Phi
,n_{sa},n_{sb})n_{sb}+\frac{\beta_{b}}{2} n_{sb}^{2}) \eqno{(12b)}$$

$$F_{L} = \frac{2Ls^{2}e^{2}}{mc} \frac{(\Phi_{0}n-
\Phi)^{2}n_{sa}^{2}n_{sb}^{2}} {(l_{a}n_{sb}+l_{b}n_{sa})^{2}}
\eqno{(12c)}$$
Here $$\alpha_{a}(\Phi ,n_{sa},n_{sb}) = \alpha_{a0} (\frac{T}{T_{ca}}-1 +
(2\pi \xi_{a}(0))^{2} \frac{(n-\Phi /\Phi_{0})^{2}n_{sb}^{2}}
{(l_{a}n_{sb}+l_{b}n_{sa})^{2}})$$
$$\alpha_{b}(\Phi ,n_{sa},n_{sb}) =
\alpha_{b0}(\frac{T} {T_{cb}}-1 +(2\pi \xi_{b}(0))^{2} \frac{(n-\Phi
/\Phi_{0})^{2}n_{sa}^{2}}{(l_{a}n_{sb}+l_{b}n_{sa})^{2}} ) $$ $\xi_{a}(0) =
(\hbar^{2}/2m\alpha_{a0})^{1/2}$; $\xi_{b}(0) =
(\hbar^{2}/2m\alpha_{b0})^{1/2}$ are the coherence lengths at T=0.

	According to the mean field approximation the transition into the
superconducting state of the section $l_{b}$ occurs at $\alpha_{b}(\Phi
,n_{sa},n_{sb}) = 0$. Because $n_{sa} \neq 0$ at $T = T_{cb}$ the
position of the superconducting transition of the $l_{b}$ section depends
on the magnetic flux value:

$$T_{cb}(\Phi) = T_{cb}[1 - (2\pi \xi_{b}(0))^{2} \frac{(n-\Phi
/\Phi_{0})^{2}n_{sa}^{2}} {(l_{a}n_{sb}+l_{b}n_{sa})^{2}}] \eqno{(12d)} $$.

At $l_{a} = 0$ the relation (12d) coincides with the relation (3) for a
homogeneous ring. A similar result ought be expected at $l_{b} \gg
l_{a}$. But at $l_{b} \ll l_{a}$ the $T_{cb}(\Phi)$ value depends strongly
on the $n_{sb}$ value. At $n_{sb} = 0$ $T_{cb}(\Phi) = T_{cb}[1 - (2\pi
\xi_{b}(0)/l_{b})^{2}(n-\Phi /\Phi_{0})^{2}]$ whereas at $l_{a}n_{sb} \gg
l_{b}n_{sa}$ $T_{cb}(\Phi) = T_{cb}[1 - (2\pi \xi_{b}(0)/l_{a})^{2}(n-\Phi
/\Phi_{0})^{2}]$. Consequently a hysteresis of the superconducting
transition ought be expected in a ring for $l_{b} \ll l_{a}$.

	To estimate the dependence of the hysteresis value on the ring
parameters, we transform the relation (12) using the relations for the
thermodynamic critical field $H_{c}=\Phi_{0}/2^{3/2}\pi \lambda_{L}\xi$;
$\alpha^{2}/2\beta = H_{c}^{2}/8\pi $ and for the London penetration depth
$\lambda_{L}=(cm/4e^{2}n_{s})^{1/2}$. We consider a ring with
$l_{a} \gg \xi_{a}(T) = \xi_{a}(0)(1-T/T_{ca})^{0.5}$. $n_{sa} \simeq
-\alpha_{a}/\beta_{a}$ in this case. Then

$$F_{GL}= F_{GLa} + Fn'_{sb}(\tau + \frac{1}{(n'_{sb}+1)^{2}} +
n'_{sb}(B+\frac{1}{(n'_{sb}+1)^ {2}} (2 +L_{I}))) \eqno{(13)} $$
Here $n'_{sb} = l_{a}n_{sb}/l_{b}n_{sa}$; $$F_{GLa} =
-sl_{a}\frac{H_{ca}^{2}}{8\pi }(1+\frac{(2\pi \xi_{a}(T))^{4}}{l_{a}^{4}}
(\frac{(n-\Phi /\Phi_{0})n'_{sb}}{(n'_{sb}+1)})^{4})$$ Because
$l_{a} \gg 2\pi \xi_{a}$, $F_{GLa} \simeq -sl_{a}H_{ca}^{2}/8\pi $.
$$F = \frac{s\xi_{a}(T)H_{ca}^{2}}{2} \frac{2\pi \xi_{a}(T)}{l_{a}}
(n-\Phi/\Phi_{0})^{2}$$ $$\tau = (\frac{T}{T_{cb}}-1)(n-\Phi
/\Phi_{0})^{-2} \frac{l_{b}^{2}}{(2\pi \xi_{b}(0))^{2}}$$ $$B_{f} =
0.5\frac{\beta_{b}}{\beta_{a}} \frac{l_{b}}{l_{a}} \frac{l_{b}^{2}}{(2\pi
\xi_{b}(0))^{2}} (n-\Phi /\Phi_{0})^{-2}$$ $$L_{I} = 4\pi
\frac{s}{\lambda_{La}^{2}} \frac{L}{l_{a}}$$ For $h > R$, $L= k4\pi
R^{2}/h$ where k = 1 at $h \gg R$. Consequently, $L_{I} = 4\pi
(l/l_{a})(lw/\lambda_{La}^{2}(T))$ in this case. At $h, w \ll R$, $L
\simeq 4l\ln(2R/w)$, therefore $L_{I} = 16\pi
(l/l_{a})(s/\lambda_{La}^{2}(T))\ln(2R/w)$ in this case.

	The numerical calculations show that the $F_{GL}(n'_{sb})$ dependence
(13) has a maximum at small enough values of $B_{f}$ and $\L_{I}$ in some
region of the $\tau$ values. The width of the $\tau$ region with the
$F_{GL}(n'_{sb})$ maximum depends on the $B_{f}$ value first of all.
At $\L_{I} \ll 2$ the maximum exists at $B_{f} < 0.4$. For example at
$B_{f} = 0.2$ and $L_{I} \ll 2$ the maximum takes place at $-1.02 < \tau <
-0.89$. This means that the transition into the superconducting state of
the section $l_{b}$ occurs at $\tau \simeq -1.02$, (that is at
$T_{cs} = T_{cb}(1 - 1.02(n-\Phi /\Phi_{0})^{2} (2\pi
\xi_{b}(0)/l_{b})^{2}$) and the transition in the normal state occurs at
$\tau \simeq -0.89$, (that is at $T_{cn} = T_{cb}(1 - 0.89(n-\Phi
/\Phi_{0})^{2} (2\pi \xi_{b}(0)/l_{b})^{2}$) if  thermal fluctuations are
not taken into account.

	The inequality $\L_{I} \ll 2$ is valid for a tube (when $h > R$) with
$2\pi lw \ll \lambda_{La}^{2}(T)$ and for a ring (when $h < R$) with $8\pi
hw \ll \lambda_{La}^{2}(T)$. The hysteresis value increases with decreasing
$B_{f}$ value and decreases with increasing $B_{f}$ value.
The $B_{f}$ value is proportional to $(n-\Phi /\Phi_{0})^{-2}$.
Consequently, the hysteresis value depends on the magnetic field value.
Because the hysteresis is absent at $B_{f} > 0.4$, it can be observed in
the regions of the magnetic field values, where $\Phi/\Phi_{0}$ differs
essentially from an integer number. The width of these regions depends on
the $0.5(\beta_{b}/\beta_{a}) (l_{b}^{3}/(2\pi \xi_{b}(0))^{2}l_{a})$ value
(see above the relation for $B_{f}$). Since $(n-\Phi /\Phi_{0})^{2} <
0.25$ and $\beta_{b} \simeq \beta_{a}$ in the real case, the hysteresis
can be observed in the ring with $l_{b}^{3} < 0.2(2\pi
\xi_{b}(0))^{2}l_{a})$. For example in the ring with $l_{b} = 2\pi
\xi_{b}(0)$ and $l_{a} = 10l_{b}$, the hysteresis can be observed at
$|n-\Phi /\Phi_{0}| > 0.35$ (if $\beta_{a} = \beta_{b}$). At $|n-\Phi
/\Phi_{0}| = 0.5$  $B_{f} = 0.2$ and the hysteresis is equal to $T_{cn} -
T_{cs} \simeq 0.03T_{cb}$ in this ring.

	\section{Transformation of thermal fluctuation energy into electric
energy}

	The hysteresis of the superconducting transition can be observed if the
maximum is high enough. The maximum height is determined by a parameter F,
which is introduced below. The hysteresis will be observed if the maximum
height is much greater than the energy of the thermal fluctuation, $k_{B}T$.
In the opposite case the thermal fluctuation switches the $l_{b}$
section from the normal state into the superconducting one and backwards
at $T \simeq T_{cb}(\Phi)$.

	Above we have used the mean field approximation which is valid when the
thermal fluctuation is small. In our case the mean field approximation is
valid if the height of the $F_{GL}-F_{GLa}$ maximum, $F_{GL,max}$, is much
greater than $k_{B}T$. This height depends on the F, $\tau$, $B_{f}$ and
$L_{I}$ values: $F_{GL,max} = FH(\tau ,B_{f},L_{I})$. The F parameter is
determined above. The $H(\tau ,B_{f},L_{I})$ dependence can be
calculated numerically from the relation (8). To estimate the validity of
the mean field approximation we ought to know the maximum value of the
$H(\tau )$ dependence: $H_{max}(B_{f},L_{I})$. We can use
the mean field approximation if $FH_{max}(B_{f},L_{I}) \gg k_{B}T$. This is
possible if the height of the ring is large enough, namely

$$h \gg \xi_{a}(0)\frac{1}{\pi H_{max}(B_{f},L_{I})}\frac{l_{a}}{w}
\frac{Gi^{1/2}} {T_{ca}/T_{cb} -1}(n- \frac{\Phi}{\Phi_{0}})^{-2}$$
Here $Gi = (k_{B}T_{ca}/\xi_{a}(0)^{3}H_{ca}^{2})^{2}$ is the Ginsburg
number of a three-dimensional superconductor. We have used the relation for
the F parameter (see above). For conventional superconductors $Gi =
10^{-11} - 10^{-5}$. $H_{max} \simeq 10^{-2}$ for typical $B_{f}$ and
$L_{I}$ values. For example in the ring with $B_{f} = 0.2$ and $L_{I} \ll
2$ the $H(\tau )$ dependence has a maximum $H_{max}(B_{f} = 0.2,L_{I} \ll
2) = 0.024$ at $\tau \simeq -0.94$. Consequently, the  value of h cannot be
very large. As an example for a ring with parameter value $B_{f} = 0.2$,
$L_{I} \ll 2$, $l_{a}/w = 20$, $T_{ca}/T_{cb} - 1 = 0.2$, and fabricated
from an extremely dirty superconductor with $Gi = 10^{-5}$, the mean field
approximation is valid at $h \gg 20 \xi_{a}(0)$ if $|n-\Phi /\Phi_{0}|
\simeq 0.5$.

	If the mean field approximation is not valid, we must take into account
the thermal fluctuations which decrease the value of the hysteresis. The
probability of the transition from normal into superconducting state and
that of the transition from superconducting into normal state are large
when the maximum value of $F_{GL}-F_{GLa}$ is no much more than $k_{B}T$.
Therefore the hysteresis can not be observed at $FH_{max}(B_{f},L_{I}) <
k_{B}T$. This inequality can be valid for a ring made by lithography and
etching methods from a thin superconducting film, where h is the film
thickness in such a ring.

	As a consequence of the thermal fluctuations, the
density $n_{sb}(r,t)$ changes with time. We can consider $n_{sb}(r,t)$ as a
function of the time only if $h,w,l_{b} \simeq \ or \ < \xi_{b}(T)$. At $T
\simeq T_{cb}(\Phi)$ (at the resistive transition) $l_{b}$ is switched
by the fluctuations from the normal state in the superconducting state and
backwards i.e. some times ($\simeq t_{n}$) $n_{sb} =0$ and some times
($\simeq t_{s}$) $n_{sb} \neq 0$. Consequently, according to the
relations (8) and (9) the direct potential difference can appear in the
region of the resistive transition of the section $l_{b}$. Thus, the energy
of the thermal fluctuations can be transformed into the electric energy of
direct current in the inhomogeneous ring at the temperature of the
resistive transition of the section with the lowest critical temperature.

	In order to evaluate the power of this transformation one ought take
into account that in the consequence of the thermal fluctuation the
Ginsburg-Landau free energy $F_{GL}$ changes in time with amplitude
$k_{B}T$. According to the relations (7), (9) and (12c) $U_{b}I/f \simeq
F_{L}$. Because $F_{L}$ is a part of $F_{GL}$ (see (12)) the power can
not exceed $k_{B}Tf$. The maximum value of the switching frequency f is
determined by the characteristic relaxation time of the superconducting
fluctuation $\tau_{GL}$: $f_{max} \simeq 1/\tau_{GL}$. In the linear
approximation region \cite{skocpol}

$$\tau_{GL} = \frac{\hbar}{8k_{B}(T-T_{c})} \eqno{(14)}$$

The width of the resistive transition of the section $l_{b}$ can be
estimated by the value $T_{cb}Gi_{b}$. $Gi_{b} = (k_{B}T/H_{c}^{2}(0)l_{b}
s)^{1/2}$ is the Ginsburg number of the section $l_{b}$. Consequently the
 power value can not be larger than

$$W = \frac{8Gi_{b}}{\hbar}(k_{B}T_{cb})^{2} \eqno{(15)}$$
and the $U_{b}$ value can not exceed

$$U_{b,max} = (\frac{8R_{b}Gi_{b}}{\hbar})^{1/2}k_{B}T_{cb} \eqno{(16)}$$

The $U_{b,max}$ value is large enough to be measured experimentally. Even
at $T_{c} = 1 \ K$ and for real values $R_{b} = 10 \Omega$ and $Gi_{b} =
0.05$, the maximum voltage is equal to $U_{b,max} \simeq 3 \mu V$. In a
ring made of a high-Tc superconductor, $U_{b,max}$ can exceed $100 \mu V$.
One ought to expect that the real $U_{b}$ value will be appreciably smaller
than $U_{b,max}$. This voltage can be determined by the periodical
dependence on the magnetic field value (see the relations (8) and (9)).

	Transformation of thermal fluctuation energy into electric energy does
not contradict the second thermodynamic law, because it is valid to within
the thermal fluctuations \cite{landau}.

\section*{ACKNOWLEDGMENTS} I thank the National Scientific Council on
"Superconductivity" of SSTP "ADPCM" (Project 95040) and the
International Association for the Promotion of Co-operation with
Scientists from New Independent States (Project INTAS-96-0452) for financial
support.

\

\end{document}